\documentclass[]{sig-alternate-05-2015}
\pdfoutput=1

\usepackage[scientific-notation=false]{siunitx}
\usepackage {graphicx}
\usepackage {color}

\usepackage{amsmath, amsfonts, amssymb}

\usepackage{balance}
\usepackage{subfig}

\usepackage{float}
\usepackage{newfloat}
\usepackage[size=small]{caption}
\usepackage{etoolbox}
\usepackage[normalem]{ulem}

\usepackage[hide]{chato-notes} 

\newcommand{\ltitle}{Cobwebs from the Past and Present: Extracting Large Social Networks using Internet Archive Data}

\begin{document}
\isbn{978-1-4503-4069-4/16/07}
\doi{http://dx.doi.org/10.1145/2911451.2911467}

\clubpenalty=10000
\widowpenalty = 10000

\title{\ltitle}

\numberofauthors{1}
\author{
\alignauthor
Miroslav Shaltev\textsuperscript{1},
Jan-Hendrik Zab\textsuperscript{1}
and\\
Philipp Kemkes\textsuperscript{1},
Stefan Siersdorfer,
Sergej Zerr\textsuperscript{2}\\
\affaddr{\textsuperscript{1}L3S Research Center, Hannover, Germany}\\
\email{\{shaltev,zab,kemkes\}@L3S.de}\\
\email{siersdorfer@outlook.de}\\
\affaddr{\textsuperscript{2}Electronics and Computer Science, University of
Southampton, Southampton, UK}\\
\email{s.zerr@soton.ac.uk} \\
}

\date{\commitDATE}

\maketitle

\begin{abstract}
Social graph construction from various sources has been of interest to
researchers due to its application 
potential and the broad range of technical challenges involved. 
The World Wide Web provides a huge amount of continuously updated data and information on a wide range of topics created by a variety of content providers, and makes the study of extracted
people networks and their temporal evolution valuable for social as well as
computer scientists. In this paper we present SocGraph - an extraction and
exploration system for social relations from the content of around 2 billion web pages collected by
the Internet Archive over the 17 years time period between 1996 and 2013. 
We describe methods for constructing large social graphs from
extracted relations and introduce an interface to study their temporal
evolution.
\end{abstract}

\section{Introduction}
\label{sec:intro}
The advances of the computer science and technology in the last decades enabled the extraction and analysis of social networks from various types of
structured and unstructured data sources. Some sources provide explicit
and easy to extract information about user relations. This includes on-line
platforms such as Facebook, Twitter, or LinkedIn that maintain user databases
and offer software interfaces for accessing contacts, friends, or followers.
However, in many cases information about social connections is hidden within
unstructured data such as Web pages and archives. In the past, personal
relationships have been extracted from textual and multimedia sources such as
books, historical repositories
\cite{Elson:2010:ESN:1858681.1858696,Bird:2006:MES:1137983.1138016,DBLP:conf/socinfo/WienekeDSLCLNPFTMNMHM13}
and web search engines
\cite{Matsuo:2006:PAS:1135777.1135837,Canaleta:2008:SES:1566899.1566939,DBLP:conf/rskt/NasutionN10}.
Despite the work towards increasing the efficiency of finding entity relations on the web
\cite{nuray2009exploiting} only recently in \cite{Siersdorfer:2015:WEL:2806416.2806582}
a method suitable for mining of large graphs has been proposed. \\
 
In this work we introduce SocGraph\footnote{http://socgraph.l3s.uni-hannover.de} - a tool for the construction and analysis of
social graphs extracted from the Internet Archive (IA) data and the exploration of
the temporal evolution of communities in a variety of applications.
\label{sec:sysarch}
\begin{figure}
 \centering
 \includegraphics[width=0.97\linewidth]{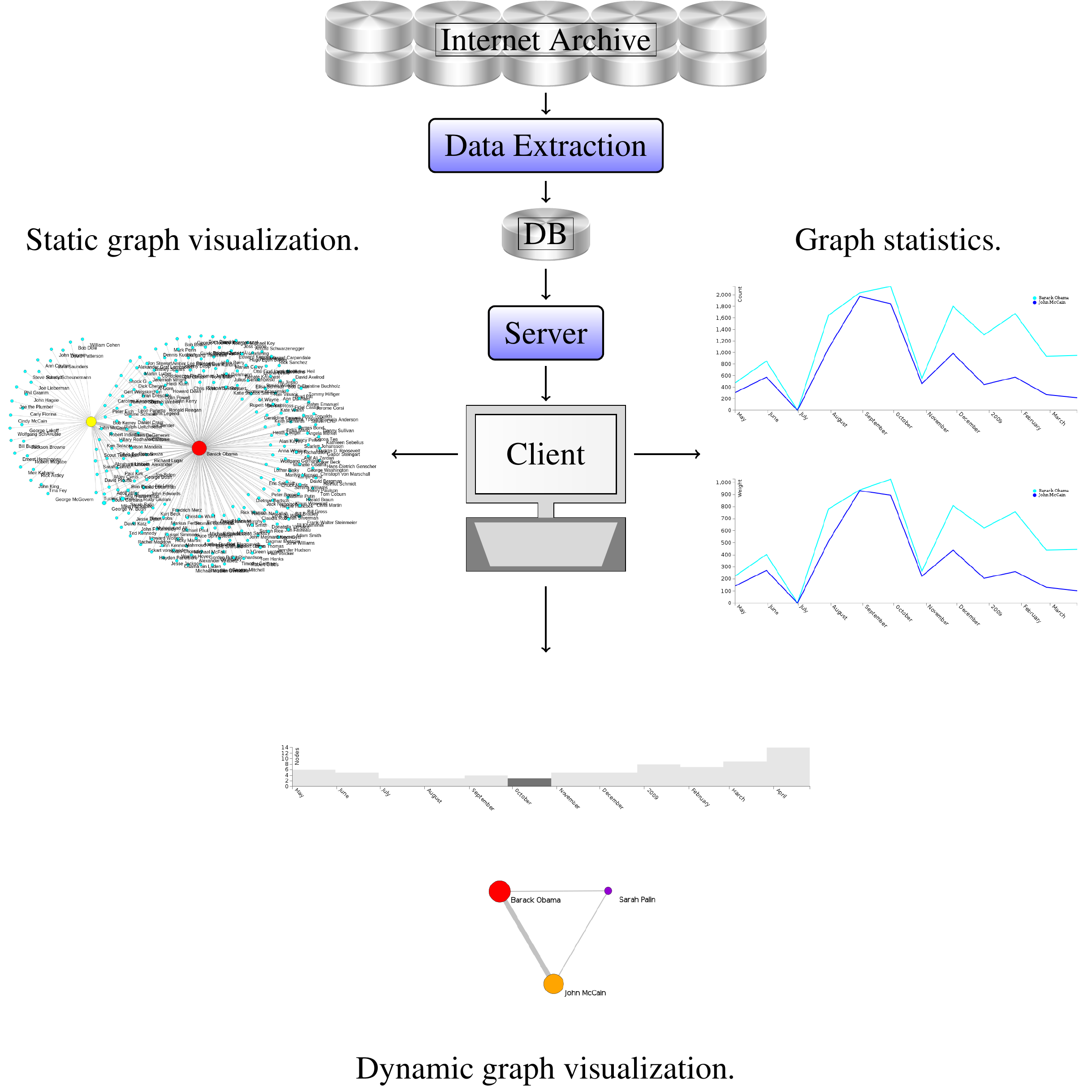}
 \caption{The overview over SocGraph system architecture and its main
 storage and visualization components.}
 \label{fig:demoarch}
\end{figure}
The objectives of the analysis include, but are not limited to shedding light on
financial and business relations on the internet, identifying ad-hoc communities
centered in social media on different events, such as solar eclipse and ''Earth
Hour'' that are concerned with artificial light pollution of the night skies of
our cities,\footnote{http://www.stars4all.eu/} or describing and
profiling of working groups in citizen science projects, where volunteers are
supporting scientists by classifying astronomic and biological phenomena in raw
data photographs.
Inspired by \cite{Siersdorfer:2015:WEL:2806416.2806582} we move a step further in that
direction and provide an interface for studying the \emph{temporal evolution} of
the social networks extracted directly from archived web page content.
For social and computer scientists our tool will provide a gateway to
the information and knowledge about connections between people stored in the
world wide web in the last few decades.

\section{System architecture}

The architecture of our system is schematically shown in Fig.~\ref{fig:demoarch}. First, we
analyze the web archive to detect co-mentions of entities in the web pages. In the
next step, at the server, we extract the temporal statistics and construct the
social graph by connecting extracted entity pairs using detected edges.\\
Finally, the user can access the
application and create, visualize, modify and interact with the graphs by
issuing new queries via a web browser based user interface. In the following we
provide an overview of the system components and show
how results are presented to the user in more detail.

\subsection{Data}
The Internet Archive (IA) is a non-profit organization crawling the World Wide Web since 1996. For
our application we have access to web pages from about $1.8\times10^{9}$
distinct URLs, collected by the IA in the time period of 17 years between
1996 and 2013. Extracting data even from a small sample of all
archived web pages is a computationally intensive and requires
parallelization of processes.

We used Hadoop and Spark
\cite{Zaharia:2010:SCC:1863103.1863113} technology on a dedicated 25 node computing
cluster with 1.3TB main memory and 268 CPU cores to extract co-mentions of persons directly from the archived documents.
We stored extracted names, patterns, URLs, date of the crawl and additional miscellaneous
information in a relational database, which can be accessed from the server
process in real time. The crawled documents of MIME type text have been encapsulated in
~346,000 Web ARChive (WARC) files.

\subsection{Extraction of Entity Pairs} 
\begin{figure}[t!]
  \subfloat[]{\label{fig:weights}
  \includegraphics[width=0.5\linewidth]{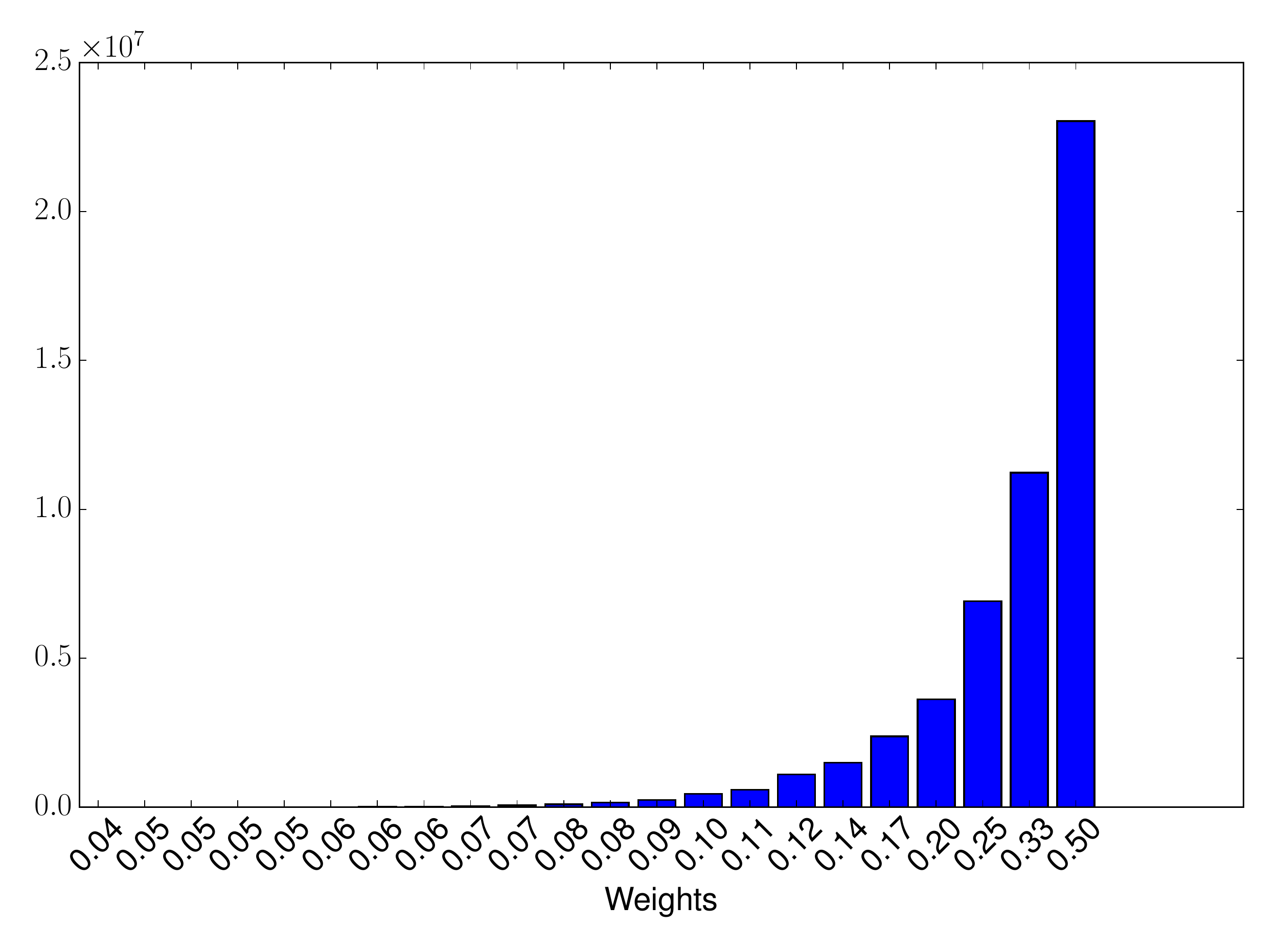}
  }
  \subfloat[]{\label{fig:edperyear}
  \includegraphics[width=0.5\linewidth]{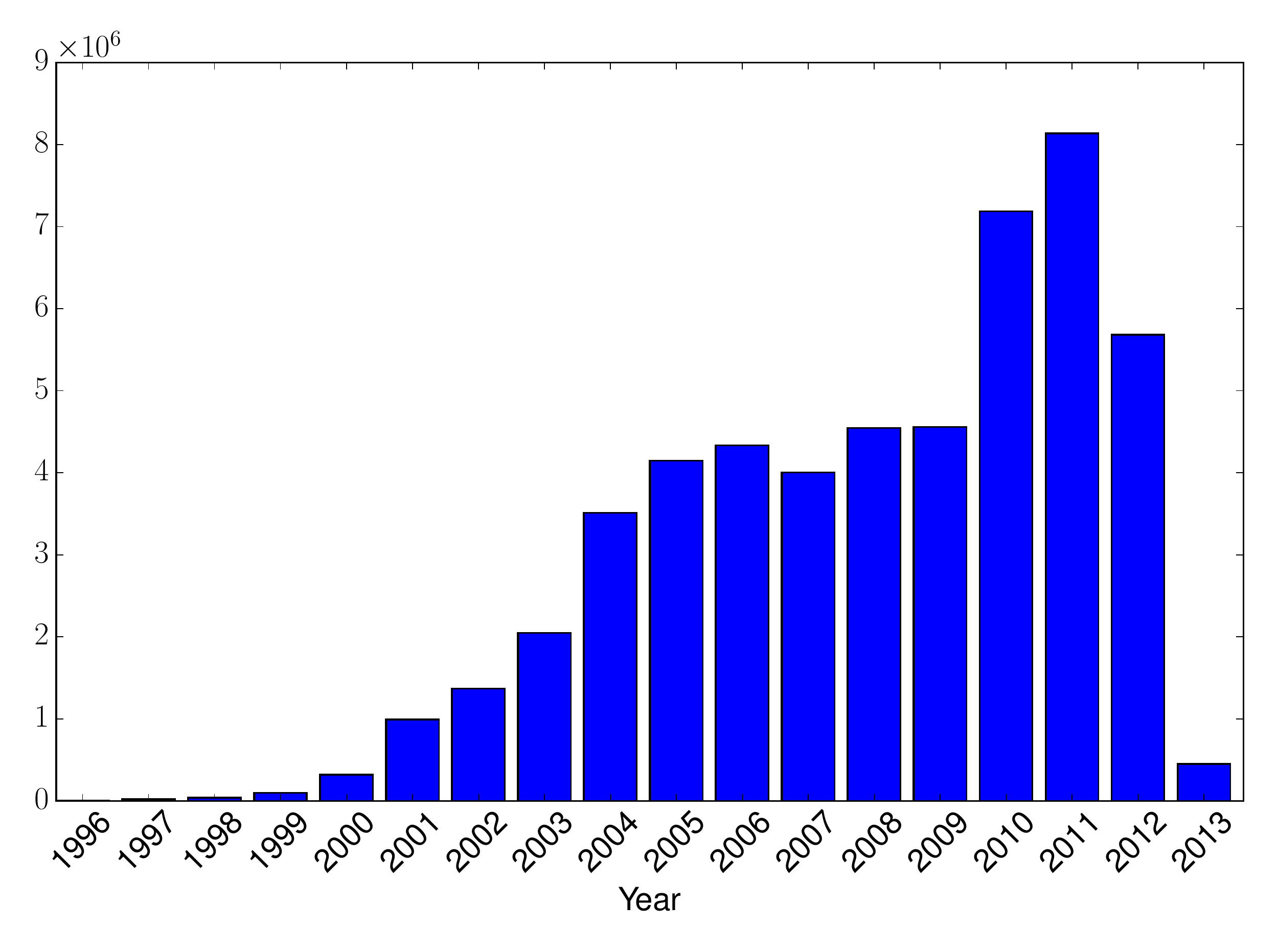}
  }
  \caption{The extracted data: (a) distribution of the weight per entry in the data set, 
  (b) number of extracted edges per year.}
\end{figure}

The entity pairs are extracted from the body of the archived documents first by
splitting the documents into sentences using the Stanford CoreNLP library \cite{manning-EtAl:2014:P14-5}.
Then extracted sentences are scanned, detecting the constructs matching the
template
\begin{equation*}
<person1><pattern><person2> 
\end{equation*}
 such as <Barack Obama><and his rival><John McCain>, using a person names dictionary and a sliding window with
a pattern length of three words. To reduce noise
in the data we exclude pairs with identical names and discard overly long sentences and patterns.
We computed the weight of the node as $1/n$, where $n$ is the number of persons
 co-occurring in the sentence. In Fig.\ref{fig:weights} we plot the distribution of
the weight per entry in the data set.
The distribution of the number of extracted pairs per year is shown in Fig.\ref{fig:edperyear}.
\begin{figure}[!ht]
 \centering
 \includegraphics[width=0.99\linewidth]{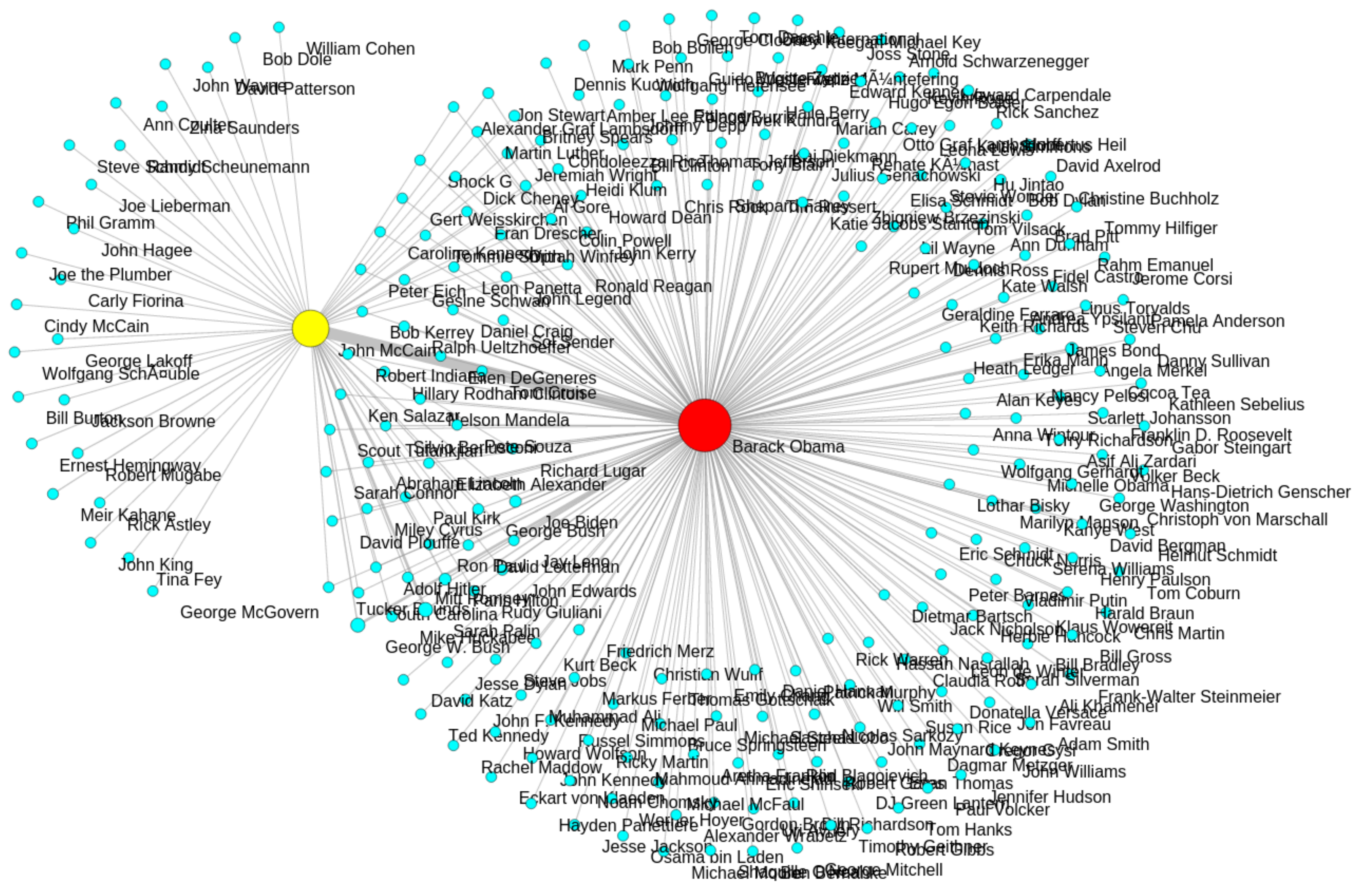}
 \vspace{10pt}
 \caption{A static graph showing the extracted social network of Barack Obama and
 John McCain for the period May, 2008 to May, 2009.}
 \label{fig:statgraph}
 \vspace{-5pt}
 \end{figure}
\subsection{Graph construction}
For building the social graph, we select extracted pairs matching the user query and user
defined parameters such as the time period $T$, and merge them into a network 
considering the node weight and the edge weight between two nodes as follows:
\begin{equation}
w=1/n\,,
\end{equation}
where $n$ is the number of co-mentioned persons in a sentence.\\\\ We then define 
node weight $\mathcal{W}_{N}$ over all considered entries $i$ as 
\begin{equation}
\mathcal{W}_{N}=\sum_i w_i 
\end{equation}
and the edge weight $\mathcal{W}_{E,kl}$ between two nodes $k$ and $l$ as 
\begin{equation}
\mathcal{W}_{E,kl}=\sum_{i} w_{k,i} + w_{l,i}\,,
\end{equation}
where $w_{k,i}$ and $w_{l,i}$ are the individual node weights.
We refer to such graphs as static graphs. To study the temporal evolution of the 
social networks we also construct dynamic graphs, consisting of sequences of 
static graphs for disjoint intervals of length $\Delta T$ within the
time period $T$.

\begin{figure*}[ht]
 \subfloat[{\small May 2008}]{\label{fig:may}
 \includegraphics[width=0.31\linewidth]{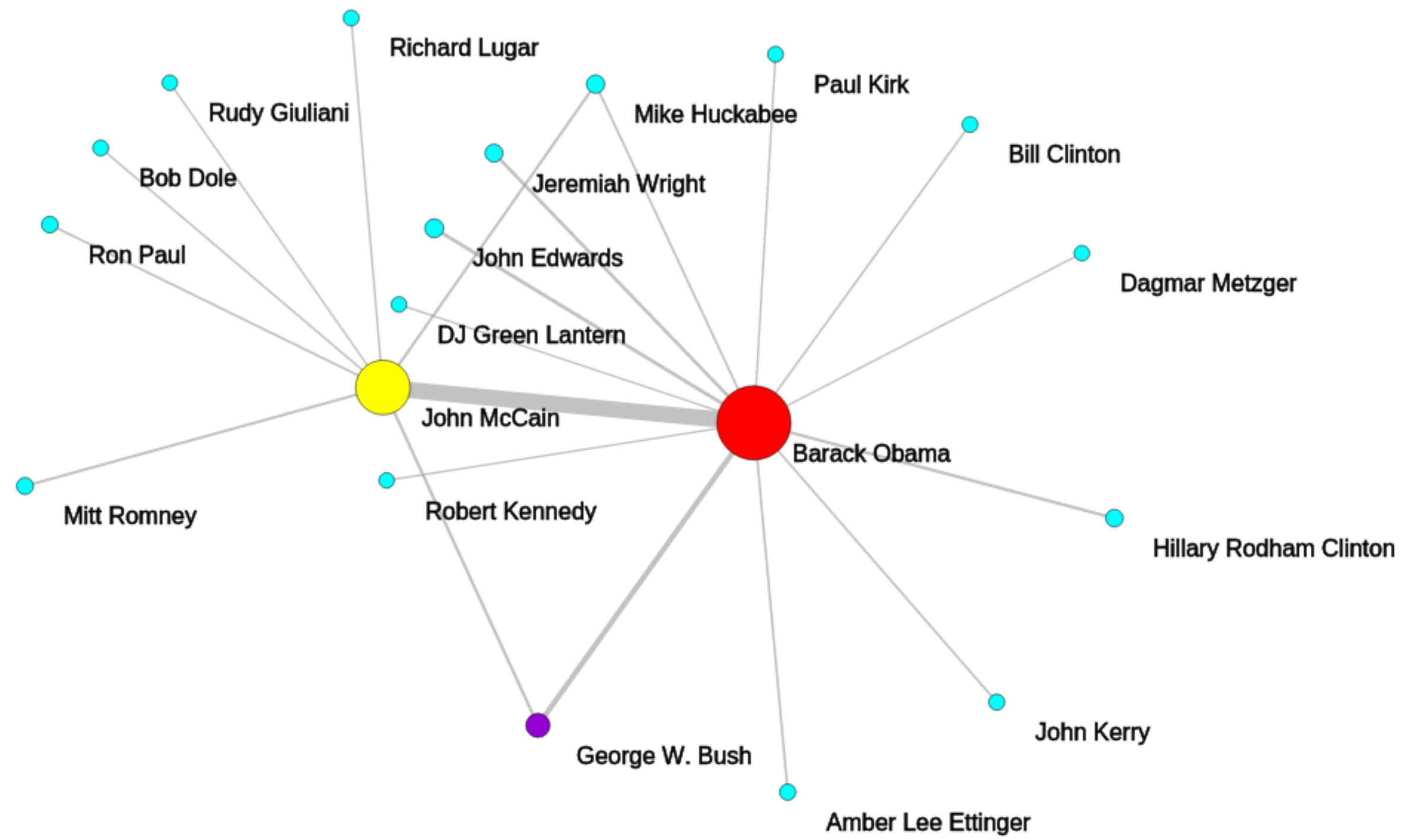}
 }
 \subfloat[{\small June 2008}]{\label{fig:June}
 \includegraphics[width=0.31\linewidth]{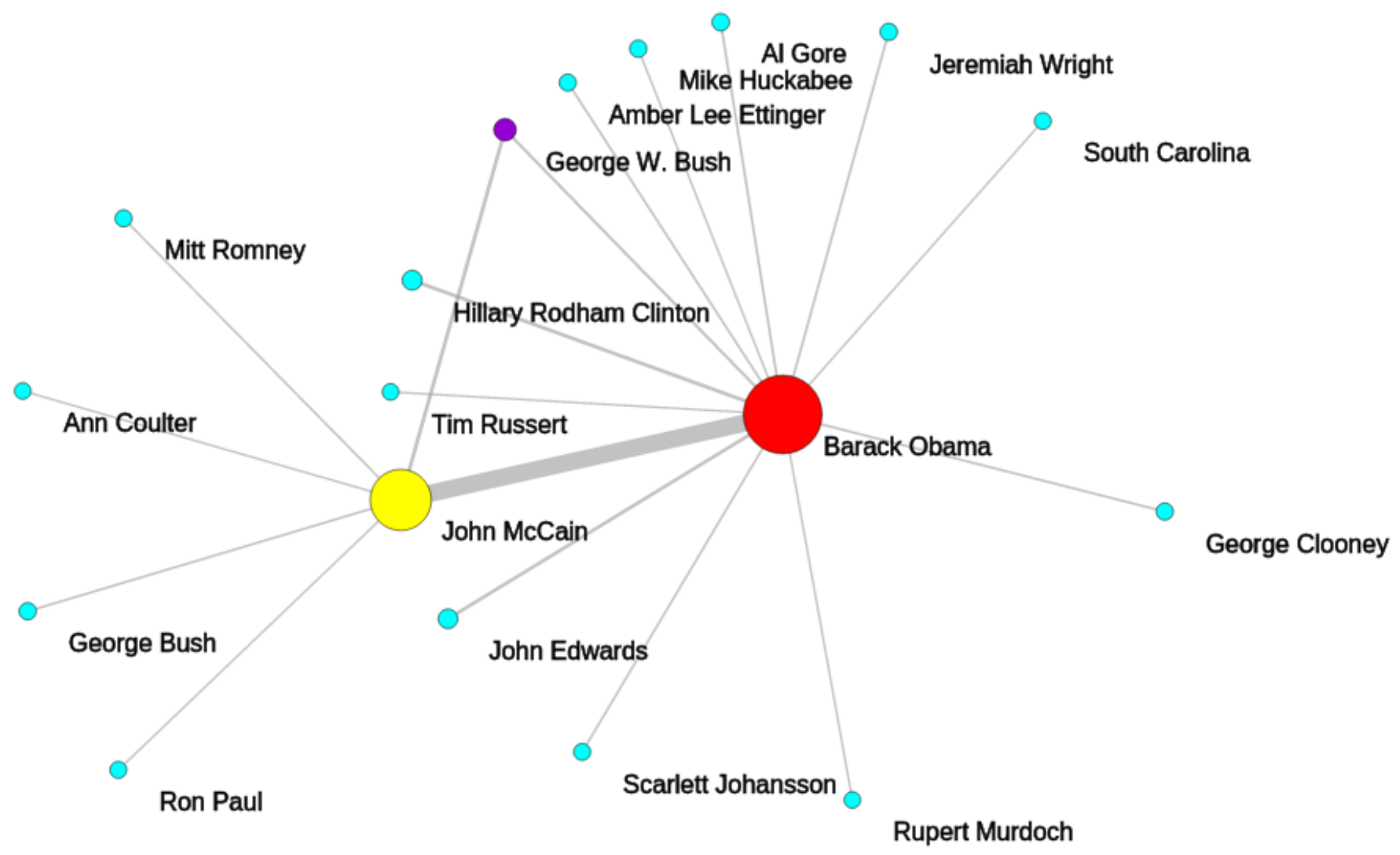}
 }
 \subfloat[{\small July 2008}]{\label{fig:July}
 \includegraphics[width=0.31\linewidth]{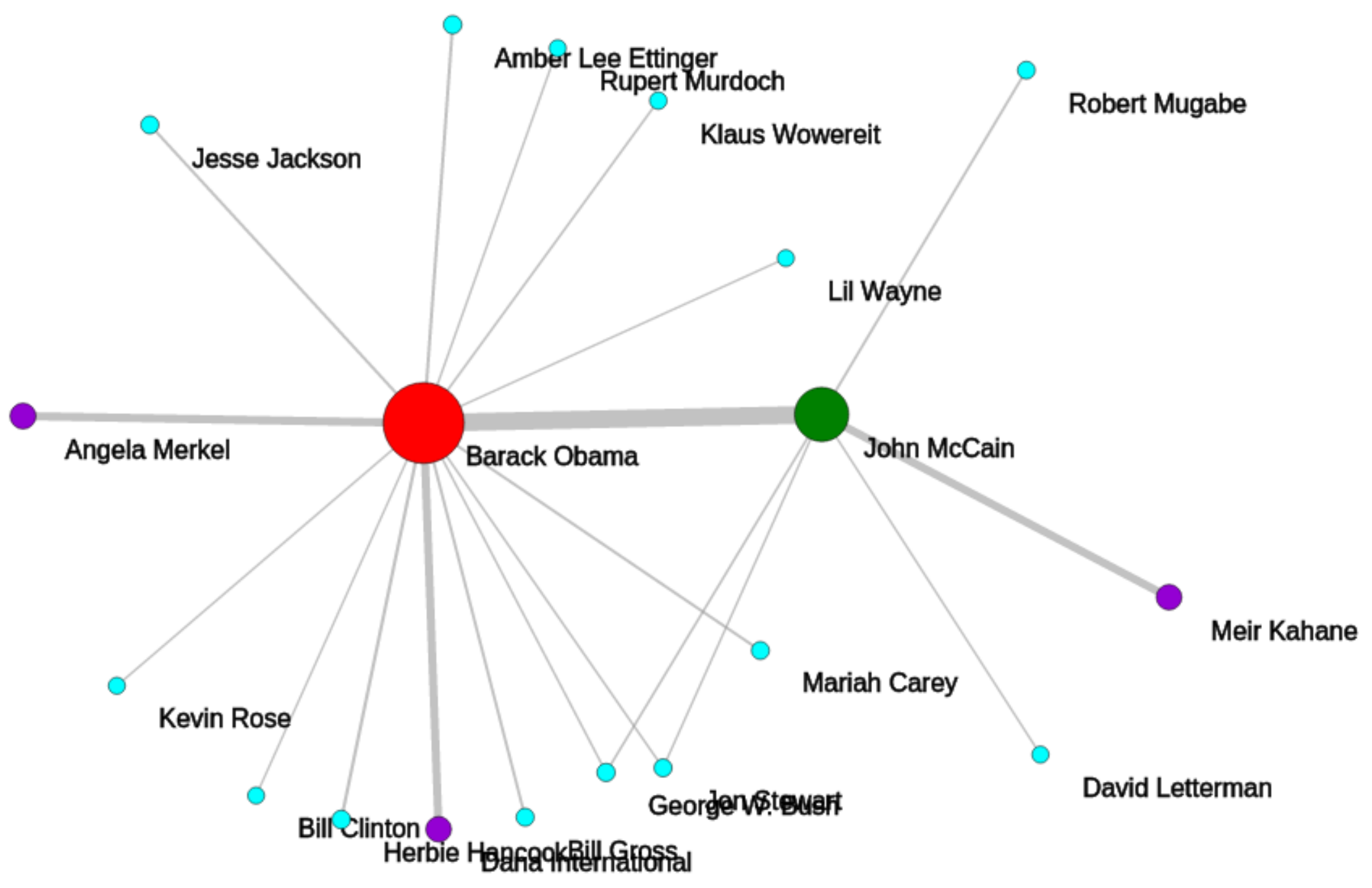}
 }\\\vspace{20pt}
 \subfloat[{\small August 2008}]{\label{fig:September}
 \includegraphics[width=0.31\linewidth]{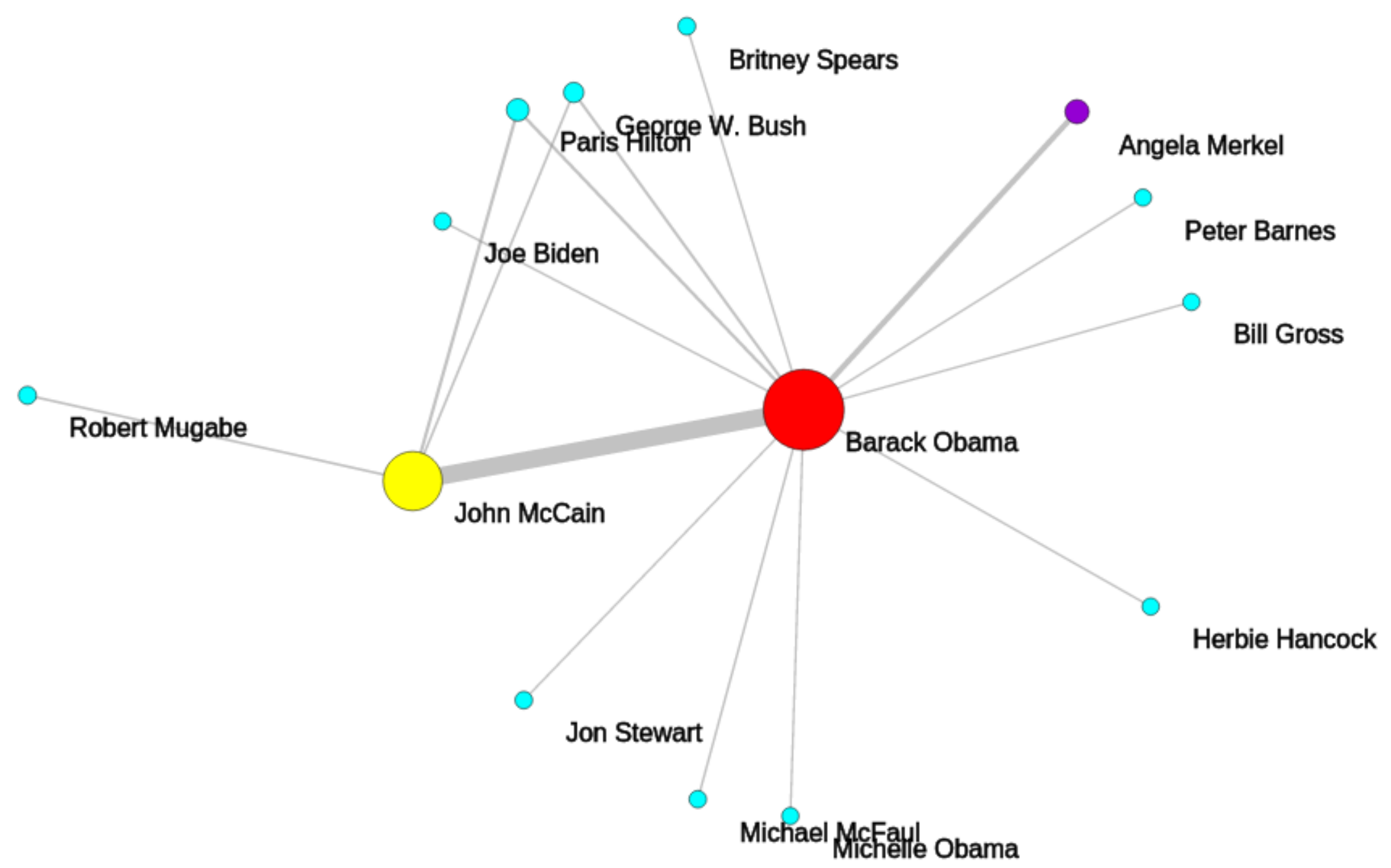}
 }
 \subfloat[{\small September 2008}]{\label{fig:November}
 \includegraphics[width=0.31\linewidth]{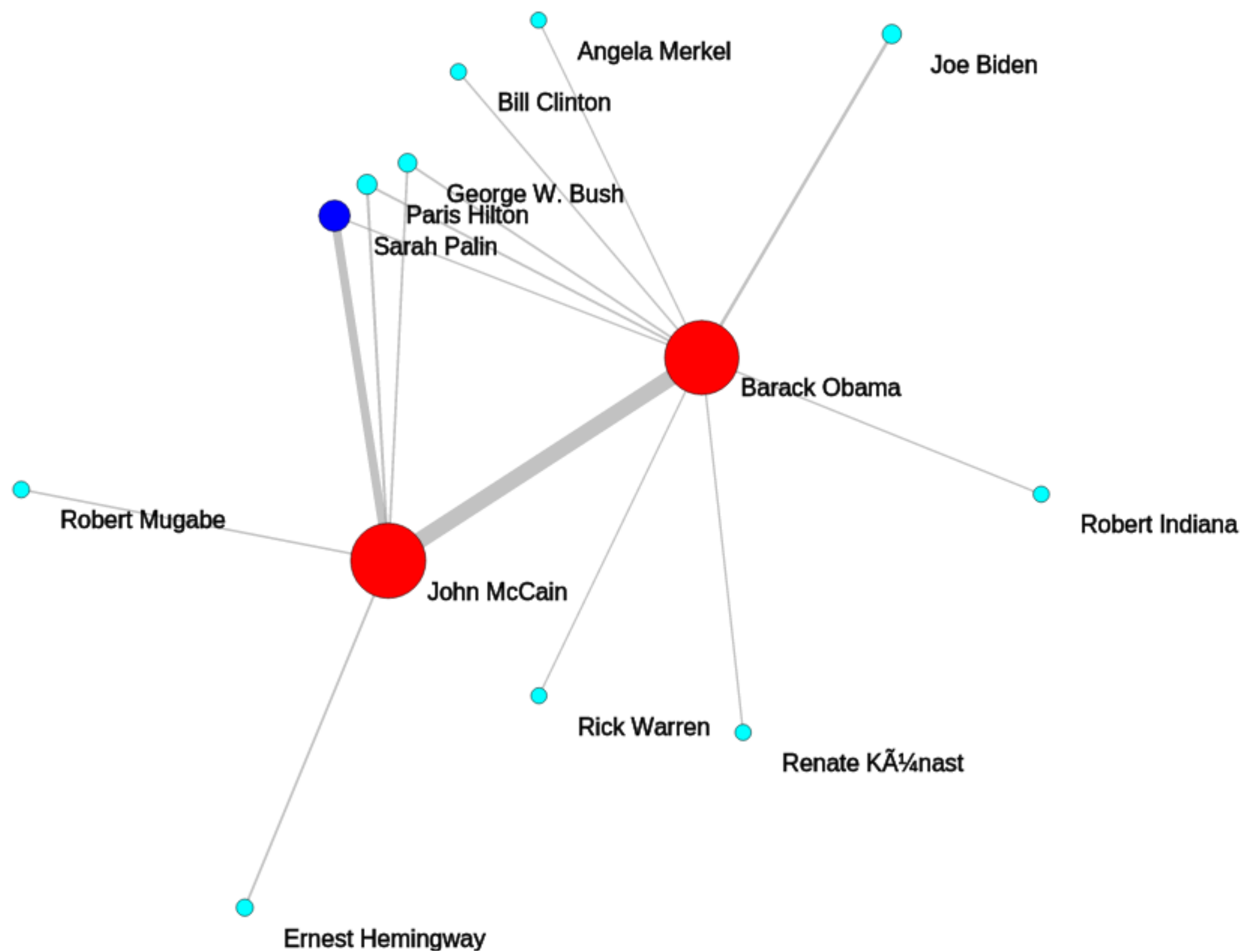}
 }
 \subfloat[{\small October 2008}]{\label{fig:December}
 \includegraphics[width=0.31\linewidth]{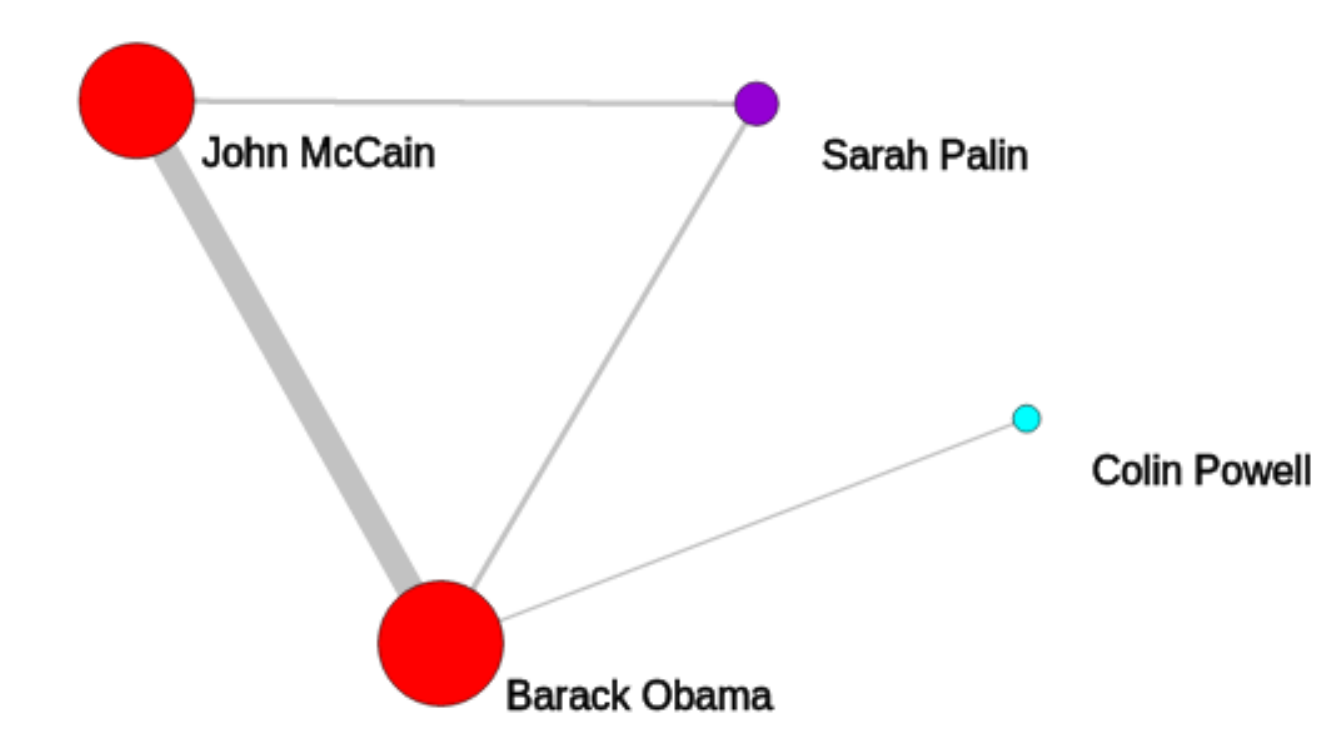}
 }\\\vspace{5pt}
 \subfloat[{\small November 2008}]{\label{fig:September}
 \includegraphics[width=0.31\linewidth]{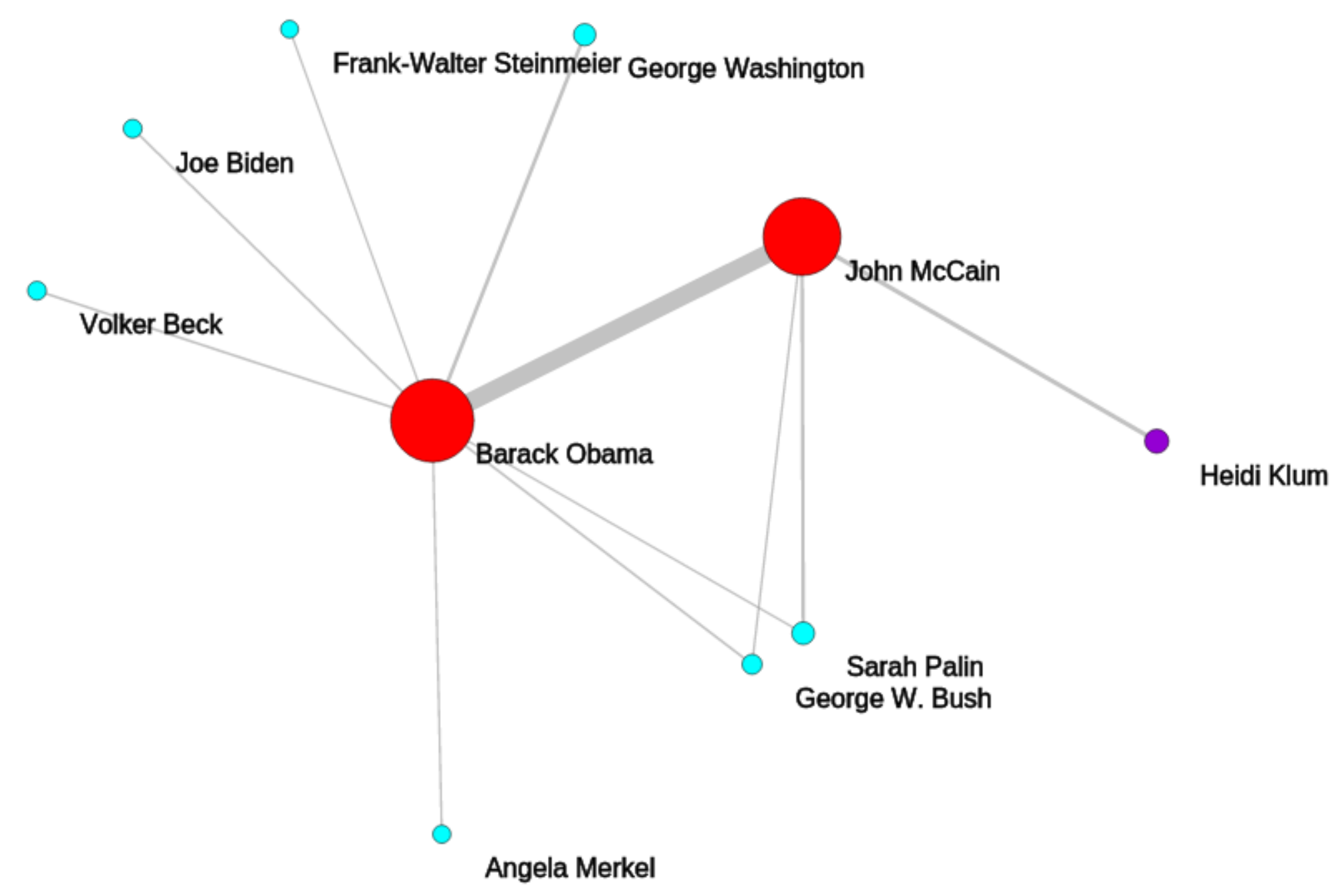}
 }
 \subfloat[{\small December 2008}]{\label{fig:November}
 \includegraphics[width=0.31\linewidth]{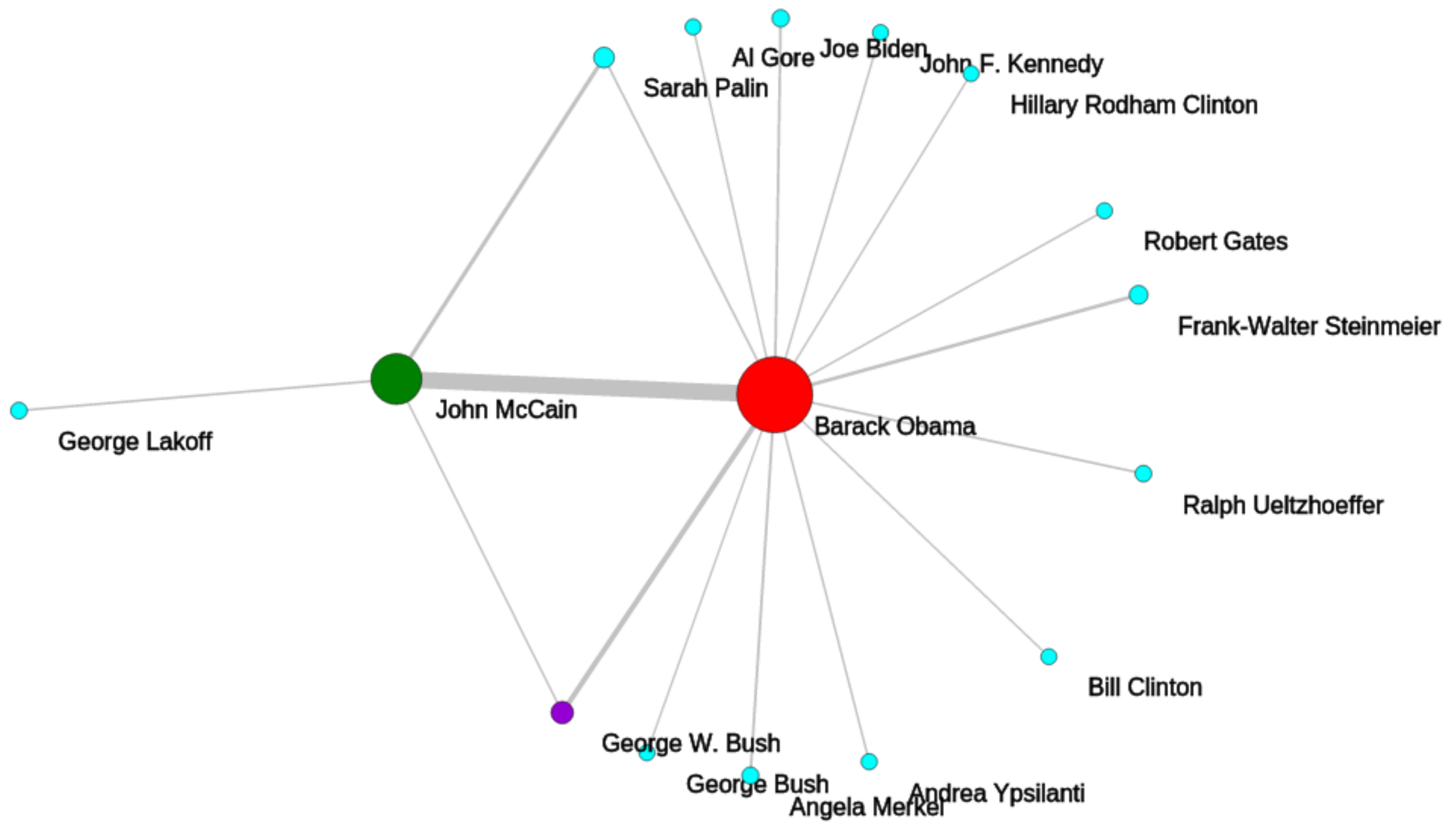}
 }
 \subfloat[{\small January 2009}]{\label{fig:December}
 \includegraphics[width=0.31\linewidth]{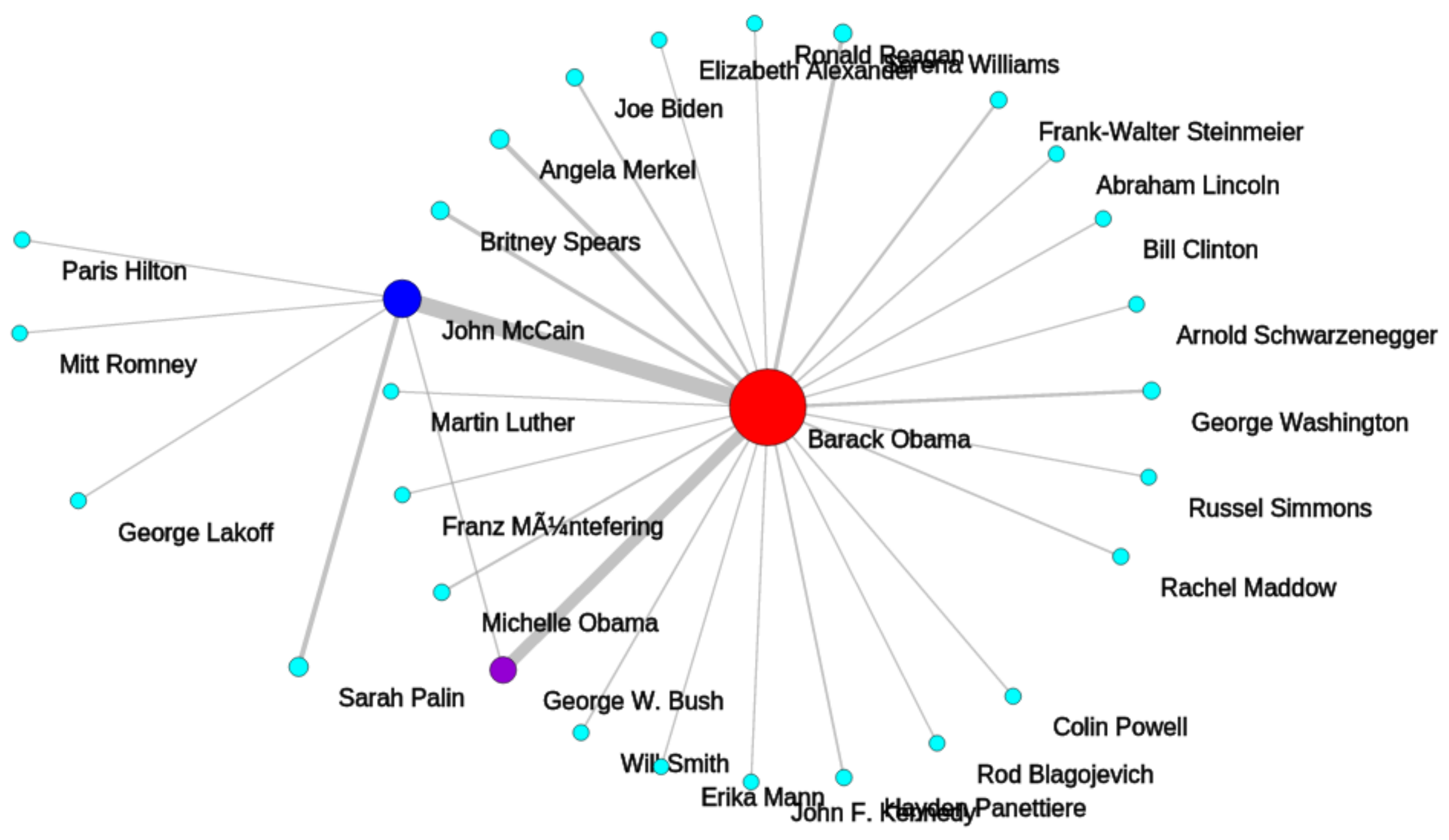}
 }\\\vspace{5pt}
 
\vspace{10pt}
 \caption{Temporal evolution of the social graph of Barack Obama and John McCain during the US election campaign 2008}
 \label{fig:dgraph}
\end{figure*}
\begin{figure}[ht]
\includegraphics[width=\linewidth]{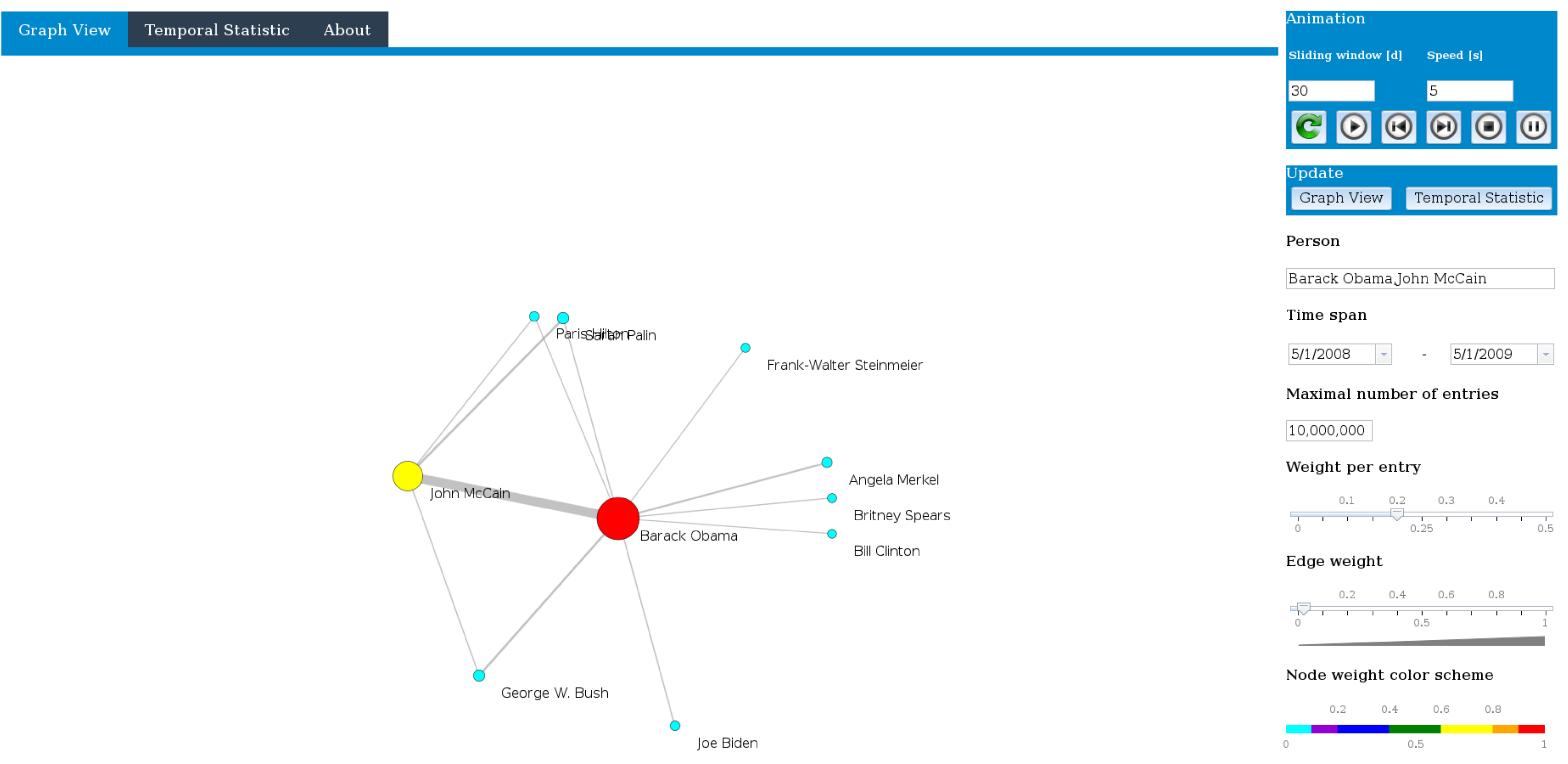}
\caption{The graphical user interface of SocGraph as shown in a web browser.}
\label{fig:system}
\end{figure}
\subsection{Graphical user interface}
The graphical user interface of SocGraph is accessible 
through a web browser, see Fig.~\ref{fig:system}. The ``Person'' input field is expecting a query such as a
name of a person or a list of person names. The ``Time
span'' selectors can be used to choose a time period of interest. 
Three filters are exposed to the user. It is possible (a) to limit the maximum
number of considered database entries through providing the value in the
''Maximal number of entries`` input field, filter the nodes (b) by their
weight through the adjustment of the ''Weight per entry`` slider and filter
the edges (c) by their weight through the ''Edge weight`` slider.\\
 The node
weights are normalized in the range from zero to one and a color scheme ranging
from blue to red is utilized to indicate particular weight value from small to
large, respectively. Overall, there are three modes of the SocGraph operation, namely:
\begin{itemize}

 \item[(i)] A static graph can be computed
and visualized over the complete dataset, providing an
overview of all co-mentions for the requested persons available in the Internet
Archive.
This mode is triggered by the ``Graph View'' button in the ``Update'' field of the control
panel on the right.
 \item[(ii)] For a specified sliding window a temporal statistical 
plot can be generated, showing the number of raw co-mentions as well 
as
the weight
of the persons of interest.
This mode is
triggered by the ``Temporal statistic'' button in the ``Update'' field of the
control panel, the results are displayed in the ``Temporal Statistic'' tab of
the demonstration interface.
\vspace*{0.5cm}

 \item[(iii)] Finally, SocGraph provides a functionality to construct and visualize 
the individual graphs for particular time periods as well as to create an
animation of graph evolution over time. The length of particular periods
(measured in days) can be entered in the ``Sliding window'' input of the ``Animation'' 
field 
of the control panel. The user obtains the dynamic graph by clicking on the
``Refresh'' button (left to the ``Play'' button) of the player control strip.\\
Once the data is loaded a time line with the number of edges in each sliding
window period appears in the ``Graph View'' tab. To start the animation, the
user should click the ``Play'' button. The animation speed can be controlled 
by the adjustment of the value (in seconds) in the ``Speed'' input. The playback
can be paused with the ``Pause button. Note, that the player also allows step by
step forward and backward rendering of the graphs, triggered by the
corresponding player control buttons.
\end{itemize}
\newpage

\section{Demonstration overview}
\label{sec:guidemo}
In the demonstration we will primarily show how the SocGraph
time travel graph system works and how the social networks are
constructed from the content of IA web pages. We will demonstrate the graphical interface usage
for static and dynamic graph visualization. Additionally, we can
elaborate in more detail on the person pair extraction process and explain
the underlying parallelization algorithms.

We will explore the social networks of Barack Obama (node
$N_{BO}$) and John McCain (node $N_{MC}$) for the one year
period from May, 2008 to May, 2009, roughly corresponding to the US presidential
election,as an example.
The weight per entry has been set to 0.2, which means that we allow data records with up to
four additional persons. The edge weight filter has been set to 0.025, in order to
not overload the graphs and focus on the interesting entities. The static graph is plotted in Fig.\ref{fig:statgraph}.
A dynamic graph has been constructed with a sliding window of 30 days. The series
of graphs are shown in Fig.\ref{fig:dgraph}. 
For instance we observe that Hillary Clinton ($N_{HC}$) node was connected to
($N_{BO}$) from the beginning and disappeared in
June 2008, corresponding to the time point, where Hillary Clinton endorsed Barack Obama and withdrew her candidacy. In following
the sizes of both nodes, $N_{BO}$ and $N_{MC}$, remain similar
until November, where the actual election took place and $N_{MC}$ drastically
reduced the weight already in following month after Barack Obama became a president of the
United States. Shortly before in September, Sarah Palins' ($N_{SP}$) talk on the
side of John McCain had positive impact on the votes for this candidate. This
fact is also reflected in our graph where $N_{SP}$ appeared in September
connected to $N_{MC}$.

\section{Discussion}
In this paper we introduced a demonstration of SocGraph - a social graph extracting 
system for large networks from Internet Archive data. In contrast to other research
concerned with graph construction from web related data, we are focused on
the temporal evolution of 
social networks implicitly contained in the stored web pages. In our future work we plan to
include pattern filtering techniques, integrate data from search engines and
evaluate event identification, as well as sentiment analysis of the
personal relationships and their evolution.
Our system will be offered as a service within the EU Project Alexandria as part of a Web Observatory accessible to social and computer scientists as
well as to general public for social network visualization and evolution
analysis. The demonstration is available on our web
page: \texttt{http://socgraph.l3s.uni-hannover.de} and can be used with
any web browser. Additionally, the web page contains a summary of the demo
applications as well as a short video tutorial.

\section{Acknowledgments}
This work is partly funded by the European Research
Council under ALEXANDRIA (ERC 339233), by the
European Commission under grant agreements 619525 (QualiMaster) and 688135
(STARS4ALL).

\bibliographystyle{abbrv}

\end{document}